\documentclass[]{mn2e}
\usepackage{epsfig}
\usepackage{amsbsy}  
\newcommand{\beq}{\begin{equation}}
\newcommand{\eeq}{\end{equation}}
\newcommand{\bea}{\begin{eqnarray}}
\newcommand{\eea}{\end{eqnarray}}
\newcommand{\Ol}{\Omega_\Lambda}
\newcommand{\Om}{\Omega_{\rm m}}
\def \om{\Omega_{\rm m}}
\def\Mp{M_{\rm Pl}}
\def \ox{\Omega_Q}

\def \P{{\rm P}}
\def \pq{p_{\rm Q}}
\def \rq{\rho_{\rm Q}}

\def \D{d_{\rm L}} 
\def \bt{{\boldsymbol{\theta}}} 
 
\def \bw{{\bf w}}

\def \lap{\stackrel{\scriptstyle \sim}{\scriptstyle <}}
\def \gap{\stackrel{\scriptstyle \sim}{\scriptstyle >}}

\begin{document}

\title[Revealing the Nature of Dark Energy Using Bayesian Evidence]
{Revealing the Nature of Dark Energy Using Bayesian Evidence}
\author[T.~D.~Saini et al.]
{T.~D.~Saini, J.~Weller and S.~L.~Bridle \\ Institute of
Astronomy, University of Cambridge, Madingley Road, Cambridge CB3
0HA.}
\date{In original form \today}
\maketitle
\begin{abstract}
We apply the Bayesian concept of `evidence' to reveal systematically
the nature of dark energy from present and future supernova luminosity
distance measurements.  We express the unknown dark energy equation of
state $w(z)$ as a low order polynomial in redshift and use evidence to
find the polynomial order, thereby establishing the minimum order
required by the data. We apply this method to the current supernova
data, and with a prior $-1 \le w(z) \le 1$ and $\Omega_m = 0.3 \pm
0.05$, obtain a large probability of $91\%$ for the cosmological
constant model, with the remaining $9\%$ assigned to the two more
complex models tested.  We also investigate the use of evidence for
future supernova data sets such as distances obtainable from surveys
like the Supernova Acceleration Probe (SNAP). Given a low uncertainty
on the present day matter density we find that, if the underlying dark
energy model is only modestly evolving, then a constant $w(z)$ fit is
sufficient. However, if the evolution of the dark energy equation of
state to linear order is larger than $|w_1| \gap 0.5$, then the
evolution can be established with statistical significance. For models
where we can assume the prior $-1 \le w(z) \le 1$, the correct
polynomial order can be established even for modestly evolving
equations of state.
\end{abstract}

\section{Introduction}
Observations of distant supernovae give strong indications that the
expansion of the universe is accelerating
\cite{Riess:98,Perlmutter:99a,Riess:01,Tonry:03}. Indications for a low matter
density universe are provided by X-ray observations of rich clusters
\cite{Bahcall:98,Mohr:99,Dodelson:00a}, which recently has been
confirmed with remarkable accuracy \cite{Allen:02}.  If cosmic
microwave background (CMB) observations \cite{Hinshaw:03} are combined
with large scale structure surveys, such as the two degree Field
Galaxy Redshift Survey (2dFGRS) \cite{Percival:02}, the best fit
cosmological model is a flat FRW universe with $\Ol \sim 0.7$ and $\Om
\sim 0.3$ \cite{Spergel:03}.

In the simplest case these observations are explained by the addition
of a cosmological constant term in the Einstein's theory of
gravity. The cosmological constant is often identified with the energy
density of the vacuum \cite{Zeldovich:68}, but explaining its small
value today ($10^{-120}\Mp^4$) in terms of fundamental physics has
remained unsuccessful \cite{Weinberg:89,Carroll:00}.  Therefore,
attempts have been made to explain the missing dark energy as the
energy density of a minimally coupled scalar field called Quintessence.

Due to the slow roll of the Quintessence field
\cite{Wetterich:88,Ratra:88,Peebles:88} the universe can become
dominated by vacuum energy and the expansion begins to accelerate.
In these models the energy in the scalar field becomes important
only at relatively late times, giving ample time for the growth of
structures in the universe; while tracker like solutions for the field
help towards ameliorating the fine tuning problem faced by a pure
cosmological constant \cite{Steinhardt:99}.

At present there are a large number of models that can describe the
observed acceleration in the expansion of the universe. These models
differ from a cosmological constant mainly through their equation of
state $\pq = w \rq$. By assuming a constant equation of state the
current bounds are $w \lap -(0.6-0.8)$ \cite{Spergel:03}. Although
most models predict $w \ge -1$, this is not necessary in non-minimally
coupled field models \cite{Uzan:99,Chiba:99a,Amendola:99} or models
with a non-canonical kinetic term \cite{Picon:00}.

One of the most promising probes for dark energy are future
supernova observations, such as the proposed Supernova Acceleration
Probe - SNAP\footnote{See at: http://snap.lbl.gov}\cite{SNAP:02}, the
w project ESSENCE\footnote{See at: http://www.ctio.noao.edu/wproject}
\cite{Garnavich:03} and the Canada - France - Hawaii Telescope Legacy
Survey (CFHTLS)\footnote{See at: http://www.cfht.hawaii.edu}. The
potential of a SNAP class experiment in distinguishing theoretical
models has been investigated in great detail in various works. 

Most methods approximate the dark energy equation of state $w(z)$ in terms of
low order polynomials, or other simple fitting functions
\cite{Efstathiou:99,Maor:01,Astier:00,Weller:01,Weller:02a,Gerke:02}.
A polynomial approximation is easily interpreted in terms of weighted
means of the true equation of state \cite{Saini:03a}. The polynomial
approximation naturally allows us to phrase three interesting
questions about the dark energy:  First is our universe described by a
cosmological constant ($w=-1$), second is the equation of state a
constant and $w
\ne -1$? and third, the most general case, is $w$ evolving with
redshift?  To answer these questions we need to decide on  
the basis of data which polynomial order is required to fit the data
\emph{adequately}, regardless of the fitted values of the parameters. 
The fact that polynomial coefficients are related in a simple way to
the true equation of state also allows asking further questions
regarding the possible time evolution of dark energy.

The Bayesian construct `evidence' answers precisely this type of
question.  The method has the usual likelihood ratio test built into
it, as well as the sensible criterion for penalising models with a
large number of parameters, known as the Occam's razor term
\cite{MacKay,Sivia}. Our main aim in this paper is to use this concept
to formulate the dark energy problem as a model selection problem, and
to quantify the requirements of the present and the future data for
elaborate modelling. In addition the method is easily extended to
differentiate between the diverse theoretical models for quintessence,
even when their parameterizations are totally unrelated.

In Section~\ref{sec:evidence} we introduce the concept of evidence and
illustrate its meaning using a simple example.
Section~\ref{sec:method} contains a brief overview of Type
Ia supernovae as standard candles, and a brief description of the SNAP mission.
In Section~\ref{sec:results} we use evidence to calculate the
acceptance probability for simple models of dark energy by using the
currently available SNe data. We investigate the future prospects
for discriminating amongst various dark energy models by simulating
SNAP like data and applying the evidence procedure. Furthermore we analyse
how prior information on the matter density and the dark energy
parameters is affecting evidence. Our conclusions are presented in
Section~\ref{sec:conclusion}.
\section{Bayesian model selection: Evidence} \label{sec:evidence}
The notion of \emph{evidence} is not frequently used in the
astronomical literature, therefore we review the Bayesian method for
model selection, and then apply it to the problem of inferring the
polynomial order required by a given data set. For further details see
Sivia (1996).

\subsection{Mathematical framework}
A model consists of a set of rules to predict data from a given set of
parameters and a {\em prior} which quantifies the probabilities of the
different parameter values in the absence of any data.  Consider a set
of models (hypotheses) $\{H\}$. From Bayes' theorem the probability that
$H$ is true is
\begin{equation}
\P(H\, |\, D) = \frac{\P(D\,|\,H) \P(H)}{\P(D)}\,\,,
\label{eq:evidence}
\end{equation}
where $D$ denotes the observed data.
This shows how our prior probability
$\P(H)$ is modified by the presence of the data
to give the {\emph{posterior} probability $\P(H\,|\, D)$.  

The
probability $\P(D\,|\,H)$ is the probability of data marginalized
over the parameter values in the model $H$.
This can be seen
more clearly from the following.  Writing the parameters for model $H$ as
${\btheta}$, Bayes' theorem gives for the posterior probability of 
the parameters given the data and model:
\begin{equation}
\P({\btheta}\,|\,D,H) = \frac{\P(D\,|\,{\btheta},H) 
\P({\btheta} | H)}{\P(D\,|\,H)}\,\,.
\end{equation}
$\P(D|{\btheta},H)$ is the usual likelihood
of data, given the model and its parameters, and $\P({\btheta}|
H)$ are the priors on the parameters. 
The required quantity is
the denominator in the right hand side
which  is found from normalizing the left hand side to unity to be
\begin{equation}
{\cal E} \equiv \P(D\,|\,H) = \int {\rm d}^n \,  \btheta
\P(D\,|\,{\btheta},H)
\P({\btheta}\,|\, H)\,\,,
\label{eq:evidencedef}
\end{equation}
where ${\cal E}$ denotes the evidence of the hypothesis $H$.

If the data is predicted by a large volume of the parameter space
allowed by the priors then the model gets a high probability
(evidence), which is a very desirable feature. The denominator in
Eqn.~(\ref{eq:evidence}) is an overall normalization constant which can
be ignored if one is interested only in the relative merit of the
various hypotheses. The term $\P(H)$ is our prior probability for the
various models being compared.  A uniform prior over all the models
being considered would express our lack of inclination for any
particular model.  In this case the posterior probability
$\P(H\,|\,D)$ in Eqn.~\ref{eq:evidence}, for the various hypotheses is
proportional to the evidence.

\subsection{Interpretation: what evidence measures}
\label{sec:gaussian}

To gain a better intuitive understanding of evidence we now carry out the
calculations analytically for a simple example.
Assume that the likelihood $P(D\,|\,\btheta,H)$ is a Gaussian about the
best fit likelihood position $\btheta_{L}$
\begin{equation}
P(D\,|\,\btheta,H)=P(D\,|\,\btheta_L,H) \,\exp \left [ -\frac{1}{2}(\btheta-\btheta_L)^T
\,{\bf F}\, (\btheta-\btheta_L) \right]
\end{equation}
where ${\bf F}$ is the usual Fisher (curvature) matrix
defined by $F_{ij}=-\partial^2 \left[ \log P(D|\btheta,H)\right] / \partial \theta_i \partial \theta_j$
evaluated at $\btheta=\btheta_L$.
Assume that the prior is also a Gaussian, but centred on
$\btheta_P$ with curvature matrix ${\bf P}$. 
Evaluating the integral in Eq.~\ref{eq:evidencedef} we find
\begin{equation}
\mathcal{E}=P(D\,|\,\btheta_L,H)\, \exp(-C) \,\left( \frac{|{\bf F} +{\bf P}|}
{|\bf P|} \right )^{-1/2}\,\,,
\end{equation}
where $C$ is a constant depending on the degree of overlap between the
prior and likelihood distributions. For the simpler case where
$\btheta_L=\btheta_P$ we find that $C=0$ and if we define volume
under the prior $V_{\rm prior} = |{\bf P} |^{-1/2} $ and the volume
under the posterior $V_{\rm post} = |{\bf F} + {\bf
P} |^{-1/2}$ then the evidence becomes 
\beq
{\cal E} \approx \P(D\,|\,{\btheta_0},H)\,\,
\frac{V_{\rm post}}{V_{\rm prior}} \,\,.
\label{eqn:anaev}
\eeq
The first term gives the usual likelihood at the best fit point, and
the second fraction is the ratio between the volume under the
posterior to the volume under the prior.  Therefore evidence takes
into account the likelihood at the best fit point, which usually
increases with the number of parameters.  However this is partially
countered by the second term --- the so called Occam's razor term ---
which ensures that models with a large number of parameters are
penalised.
\section{Supernovae}
\label{sec:method}
Type Ia supernova appear to be an excellent standard candle
\cite{Perlmutter:97,Riess:98}, with a small dispersion in 
apparent magnitude, $\sigma_{\rm mag} = 0.15$, with no evidence for 
evolution with redshift. The apparent magnitude is related
to the luminosity distance through
\bea
 m(z) ={\cal M} + 5\,\log D_{\rm L}(z)
\label{eq:mz}
\eea
where ${\cal M}=M_0 + 5\log \left[(c/H_0)/{\rm Mpc}\right] + 25$.  The
quantity $M_0$ is the absolute magnitude of Type Ia SNe and $D_{\rm L}(z) =
\D(z)/(c/H_0)$ is the Hubble constant free luminosity
distance. The combination of the absolute magnitude and the Hubble
constant, $\cal M$, can be calibrated by low redshift supernovae, for
instance the C\'alan/Tololo sample
\cite{Hamuy:93,Perlmutter:99a}. The
dispersion in the magnitude, $\sigma_{\rm mag}$, is related to the
uncertainty in the inferred distance, $\sigma$ by
\begin{equation}
\frac{\sigma}{\D(z)} = \frac{\ln 10}{5} \sigma_{\rm mag}\,\,. 
\end{equation} 
This is about $7\%$ for $\sigma_{\rm mag} = 0.15$.  In our simulations
we assume that the errors in the luminosity distance are Gaussian. We
neglect systematic errors in our calculations.  

The SuperNova Acceleration Probe (SNAP) survey is expected to observe
about $2000$ Type Ia supernovae up to a redshift $z \sim 1.7$,
each year (Aldering et al.\ 2002). Although the expected distribution
of SNe is complicated, for our calculations we assume them to be
uniformly distributed.  Since a single supernova measures the
luminosity distance with a relative error of $\sim 7\%$, binning the
supernovae in $\sim 50$ bins would give a relative error in the
luminosity distance of about $\sim 1\%$.

For our simulations we calculate the luminosity distance in $50$
redshift intervals up to a maximum redshift $z=1.7$. We assume a
relative error of $1\%$ in the luminosity distance. Note that we
\emph{do not} add noise to the simulated distances. Therefore, our
results only give the ensemble average of the various quantities that
we quote below.

To decide which polynomial order is best suited to the data we now
apply the notion of evidence. To label different models we choose the
letter $N$ --- the order of the polynomial approximation. The case of
cosmological constant is denoted as $N=-1$. The only free parameters
are the matter density $\om$ and the polynomial coefficients
$\bw=(w_i)$. In reality we also have to marginalize over $\cal M$, but
for simplicity we ignore this in our calculations. Therefore, our free
parameters are $\bt=(\om,\bw)$.  

The likelihood function $\P(D\,|\,{\btheta},N)$ is given by
\begin{equation}
P(D|\bt,N) = {\cal N} \prod_{i=1}^{N_{\rm dat}} \exp \left [
-\frac{1}{2} \left ( \frac{ \D^{\rm fit}(z_i,\bt)-\D(z_i)}{\sigma_i}
\right )^2 \right ]\,\,, 
\label{eq:lhood}
\end{equation}
where the index $i$ ranges from $1$ to $N_{\rm dat}$, the number of
supernovae in our sample (or the number of bins in redshift) The data
is described by the measured luminosity distance $\D(z_i)$, the
dispersion $\sigma_i$ and the redshift $z_i$. Also 
\beq
\begin{array}{l}
\displaystyle
\D^{\rm fit}(z) =  \displaystyle \frac{c(1+z)}{H_0} \times \\
 \\
 \displaystyle \int_0^z \frac{(1+z^\prime )^{-3/2}\,{\rm d} z^\prime}
{\sqrt{\om +  \ox \exp\left\{3\int_0^{z^\prime} w(z^{\prime
\prime})/(1+z^{\prime \prime}) {\rm d} z^{\prime \prime}\right\} }}\,
,
\end{array}
\label{eq:lumdist}
\eeq
where $w(z) = \sum_{i=0}^N w_iz^i$, $\om$ is the present day
fractional energy density in pressure-less matter and $\ox=1-\om$ is
the present day fractional density in dark energy. In
Eqn.~\ref{eq:lhood} the normalization constant ${\cal N}=
{\left(2\pi\right)^{-N_{\rm dat}/2}}/{\prod_{i=1}^{N_{\rm
dat}}\sigma_{i}}$.

\section{Evidence analysis} 
\label{sec:results}

We begin this section by comparing evidence values for different
theoretical models fitted to the current supernova data. We then
repeat this exercise by simulating future data for various assumed
input models and show how well evidence picks out the correct model.
We then investigate the issues related to the effect of uncertainty in
the matter density, and the effect of different priors on the dark
energy parameters.

\subsection{Current supernova data}

Although the quality of data will continue to improve, it is
instructive to see what the current supernova data tells us about which
model to trust. For the analysis we use the compilation of supernova
luminosity distances and redshifts by Tonry et al. (2003).  We apply a
prior of $\Om = 0.3\pm 0.05$ (similar to Allen, Schmidt \& Fabian
2002).
\begin{figure}
\epsfig{file=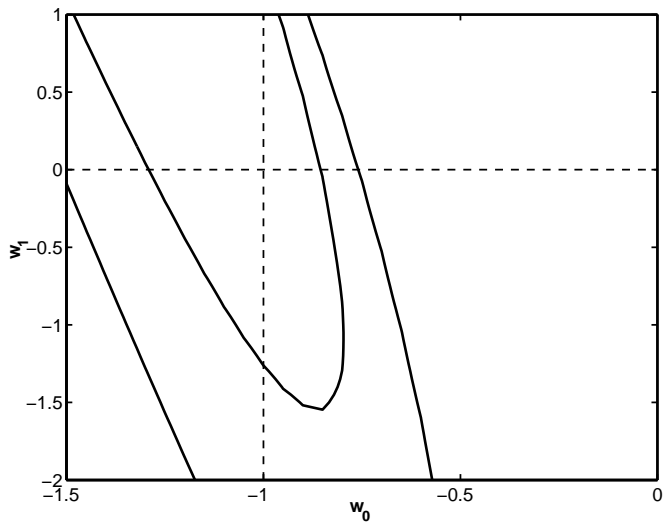,height=7cm}
\caption{The 68\% and 95\% joint likelihood contours for $w_0 - w_1$
with current Supernova data {\protect\cite{Tonry:03}}, marginalized
over $\Om$ with a Gaussian prior, $\Om=0.3\pm 0.05$, and a uniform
prior for the equation of state in the range $-1.5 \le w_0 \le 0$ and
$-2.0 \le w_1 \le 1.0$.  }
\label{fig:current}
\end{figure}
In Fig.~\ref{fig:current} we plot the $68\%$ and $95\%$ contours.  As
expected the error bars are large. The peak of the likelihood surface
lies in the region where $w<-1$.  Table~\ref{tab:evidence} gives the
computed evidence values for various models with two different priors
for $w(z)$: the first row gives the evidence values with the prior
$-1.5 \le w_0 \le 0$,  $-2 \le w_1
\le 1$ (limits of Figure~\ref{fig:current}), and  the second row
gives evidence with the prior $-1 \le w(z) \le 1$, in the range $0 \le
z \le 1.7$. As expected from Figure~\ref{fig:current} the simplest
$\Lambda$ model has the largest evidence with either prior. The data
favours the simplest model but still allows the possibility of some
evolution in agreement with the contours plotted in
Fig.~\ref{fig:current}. Since the peak of the posterior does not lie
in the region given by the second prior $-1
\le w(z) \le 1$, we find that with this prior the data is more
consistent with the cosmological constant.

\begin{table}
\begin{center}
\begin{tabular}{c|c|c|c|}
\hline\hline
{\rm Prior} & $w = -1 $ & $w=w_0$ & $w=w_0+w_1z$ \\
\hline
$-1.5\le w_0\le1$ & \raisebox{-1.5ex}[0pt]{$66.6\%$} &  \raisebox{-1.5ex}[0pt]{$18.1\%$}  & \raisebox{-1.5ex}[0pt]{$ 15.3\%$}\\	
$ -2\le w_1 \le 1$ & & & \\
& & & \\
$-1\le w(z)\le 1$ & $91.0\%$ &  $6.0\%$  & $ 3.0\%$\\	
\hline
\hline
\end{tabular}

\caption{Evidence for the Supernova
luminosity distances given in Tonry et al. (2003). We have used the
tight prior $\Omega_m = 0.3 \pm 0.05 $ for this calculation.}
\end{center}
\label{tab:evidence}
\end{table}

\subsection{The discriminatory power of future supernova surveys}
We simulate luminosity distance for a SNAP like experiment as
described in Section~\ref{sec:method}.  First we consider a linearly
evolving equation of state given by $w(z) = -0.8 + 0.6z$ with
$\om=0.3$. This choice of parameters illustrates certain degeneracies
first described in Maor et al. (2002)
In Fig~\ref{fig:likesnap} we plot the $68\%$ confidence contours in
the $w_0$--$w_1$ plane, marginalized over $\Om$, for different priors
on $\Om$. For the dark energy we chose the priors $-1.5 \le w_0 \le 0$
and $-2 \le w_1 \le 1$ in this calculation.  We address the issue of
priors on the dark energy in detail in Section~\ref{sec:priors}. We see
how the joint likelihood contours tighten as $\Om$ is increasingly
constrained.   

We also plot contours using a Gaussian approximation to the peak of
the posterior for a prior $ 0 \le \om \le 1$.  Comparing with the
corresponding exact calculation shows that the Fisher matrix error
bars are very misleading. This is not surprising since Fisher matrix
only gives the minimum variance for the extracted parameters. The
correct contours also show that the distribution of the equation of
state parameters $w_0$ and $w_1$ is not close to Gaussian. Therefore,
for our calculations below we do not use Fisher matrix approximations
described in Sec.~\ref{sec:gaussian}.

We now investigate how well the evidence can infer the required
polynomial order for the equation of state. For this purpose we choose
to consider fitting models only up to a linear order, including the
case of a cosmological constant model ($\Lambda$ model).  Therefore,
our model space comprises (1) the $\Lambda$ model with $w(z) = -1$;
(2) the constant equation of state model with $w(z) = w_0$ and (3) the
linearly evolving model with $w(z) = w_0 + w_1z$.
\begin{figure}
\epsfig{file=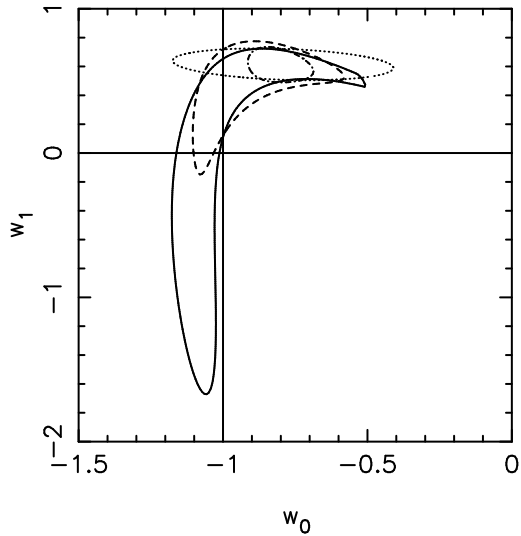,height=7cm}
\caption{The marginalized 68\% joint likelihood contours in
the $w_0-w_1$ plane for an input model $w(z)= -0.8+0.6z$ and $\Om=0.3$
for various priors on $\Omega_m$. The solid line is for a flat prior in
the range $0<\Om<1$, the dashed line a Gaussian prior with $\Om=0.3\pm
0.1$ and the dot-dashed line a Gaussian prior with $\Om=0.3 \pm
0.05$. The dotted line is the result of a Fisher matrix analysis.}
\label{fig:likesnap}
\end{figure}
In Figure \ref{fig:ev} we plot the normalized probabilities as a
function of fitting models, for various input models. For this figure
we assumed a tight prior on $\Om = 0.3
\pm 0.05$.  Evidence clearly picks out the correct polynomial order
for the $\Lambda$ model (solid line), the model with $w(z) = -0.7$
(short dashed line) the model with $w(z) = -0.8+0.6z$ (dotted
line). Here we have also considered a model with a weaker redshift
evolution, $w(z)=-0.8+0.3z$ (long-dashed line). We find that for this
case the evidence marginally prefers a model with no redshift
evolution, although
the evidence is very similar for all three models.  The evolution in
the equation of state in this model is clearly not sufficient to be
established by SNAP. In general the evidence tends to be conservative
in this respect and favours simpler models.
\begin{figure}
\epsfig{file=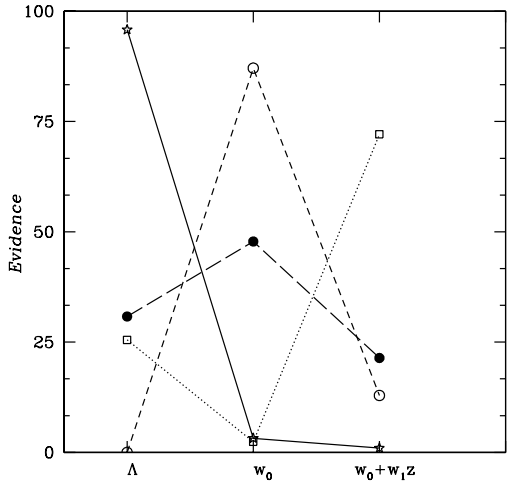,height=7cm}
\caption{Evidence values (in percent) for various models, and various 
levels of fit as indicated on the x-axis, with a Gaussian prior
$\Om=0.3 \pm 0.05$. The solid line is for an input  $\Lambda$ model, 
the short dashed line for an input  model with $w(z)=-0.7$, the
long-dashed line for a model with $w(z) = -0.8+0.3z$ and the dotted
line for $w(z) = -0.8+0.6z$.  }
\label{fig:ev}
\end{figure}

\subsection{Effect of uncertainty in $\om$}

The greatest uncertainty in constraining $w(z)$ comes from the fact
that the value of $\om$ is not very well known. To illustrate the
effect of choosing different priors for $\Om$ we calculate the
evidence for model $w(z) = -0.8 + 0.6z$ with the following three
priors on $\om$: (1) A uniform prior in the range $0< \Om < 1$, (2) a
Gaussian prior with $\sigma_{\om} = 0.1$ and (3) a Gaussian prior with
$\sigma_{\om} = 0.05$.
\begin{figure}
\epsfig{file=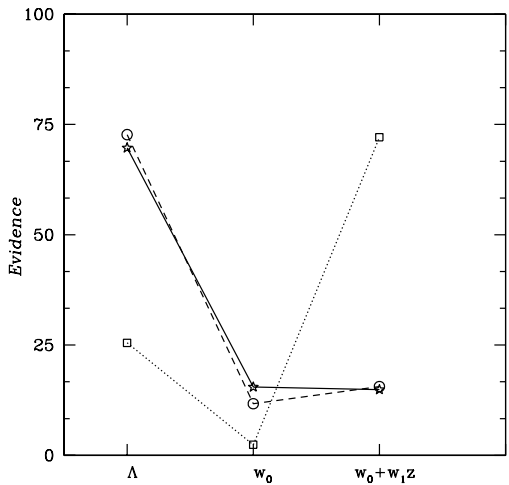,height=7cm}
\caption{Evidence values for the model $w(z)=-0.8+0.6z$ with various levels of fit
as indicated on the x-axis. The solid line is for a uniform prior in the range
$0<\Om<1$, the dashed line for a Gaussian prior, $\Om = 0.3\pm 0.1$ and
the dotted line as in Fig.~\ref{fig:ev} a prior of $\Om=0.3\pm0.05$.
}
\label{fig:ev_om}
\end{figure}
Our results are plotted in Fig.~\ref{fig:ev_om}. The correct model is
picked out only in the case where the prior on the matter density is
the tightest, $\sigma_{\om}= 0.05$.  

The real difficulty is the degeneracy that exists between a fast
evolving model and a cosmological constant model with an incorrect
value of $\om$
\cite{Maor:02}.  This can be understood from Fig.~\ref{fig:surp1},
where we plot the log-evidence for the three models as a function of
assumed values of $\om$.  We see that the evidence for the true linear
model is nearly constant and stays above the evidence for the other
two models until $\om \sim 0.46$, where a degenerate case appears
which fits best to the cosmological constant model, and the worst to
the linear model.  In fact, the $\Lambda$ model with $\om=0.46$ and
the linear model with the correct $\om$ have nearly the same
likelihood, so the evidence decides on the basis of the Occam's razor
term and penalises the more complex model. As a result the uniform
prior of $0<\Om<1$ prefers the $\Lambda$ model, and since the Gaussian
prior with $\om = 0.3 \pm 0.1$
model
does not rule out $\om=0.46$ sufficiently strongly,
it still assigns the largest probability to the $\Lambda$
model. In fact the correct inference is drawn if either the priors
on $\om$ are tight, or if we have a hard upper bound at $\om \sim 0.4$,
as Fig.~\ref{fig:surp1} clearly demonstrates.

This result can also be understood in terms of Fig.~\ref{fig:likesnap}.
The cosmological constant model corresponds to the intersection
of the axes at $w_0=-1$, $w_1=0$. 
For the two widest $\om$ priors the 68 per cent contours
almost include the cosmological constant model, which explains
why evidence uses Occam's razor to pick this model.
Whereas for the tightest prior on $\om$ the 68 per cent confidence
contours are far from the cosmological constant model, and 
indeed, as expected, the integrated probability (evidence) disfavours it.

Constraints on $\om$ can be obtained by including other cosmological
measurements with the supernova data.  Some probes such as the CMB and
cosmic shear (see Refregier et al, 2003) are sensitive to a
combination of $w$ and $\om$, so this must be taken into account.  In
addition care must be taken since the CMB probes the equation of state
at a higher mean redshift and so may have a different effective
constant equation of state (eg. Saini et al. 2003) and is sensitive to
perturbations. On the other hand there are some local measures that
are largely insensitive to $w$ or its evolution, such as the baryon
fraction in clusters and mass-to-light ratios~\cite{bahcallcdoy00}.  A
recent analysis of the baryon fraction in clusters using Chandra data
\cite{Allen:02} puts a ten per cent uncertainty on the matter density
when HST key project and nucleosynthesis information is used.

\begin{figure}
\vbox{\center{\centerline{
\mbox{\epsfig{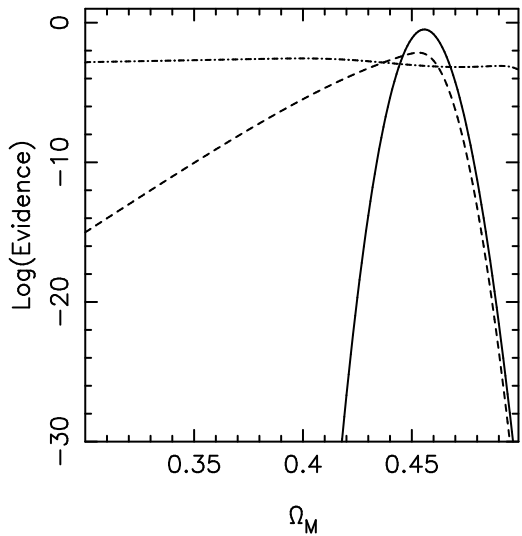}}
}}}
\caption{
Log of evidence plotted for the input model $w_0=-0.8$ and $w_1=0.6$
with $\om=0.3$ as a function of $\om$ assumed for the fit. The solid
line corresponds to the fit to a cosmological constant, the dashed
line corresponds to the fit to a constant $w$ and the dot-dashed line
corresponds to the linear fit respectively.  It should be noted that
the evidence is left un-normalized.}
\label{fig:surp1}
\end{figure}

\subsection{Effect of priors on w}
\label{sec:priors}

The evidence is affected by the assumed choice of priors on the
parameter values. Qualitatively, increasing the width of the priors on
$w$ will increasingly disfavour models with more parameters, which is
clearly seen in Eqn.~\ref{eqn:anaev}. The priors in the previous
section were arbitrarily chosen to be consistent with the input models
and to have enough room for a fair degree of uncertainty prior to the
data. We also note that although in principle one could imagine
physical reasons for limiting $w_0$ within a certain range, the
dimensional quantity $w_1 = dw/dz$ could in principle be arbitrarily
large.  Therefore, a better, physically motivated prior is not on the
polynomial coefficients but on the $w(z)$ itself. For a subclass of
dark energy models the equation of state satisfies the constraint $-1
\le w(z) \le 1$.  Figure~\ref{fig:evnew} shows the effect of choosing
this prior.  We find that the evidence is better able to pick the
correct model in this case relative to the choice of priors in the
last section.  For this subclass of theoretical models the prospects
for disentangling the dark energy properties are seen to be much
better.  

\begin{figure}
\epsfig{file=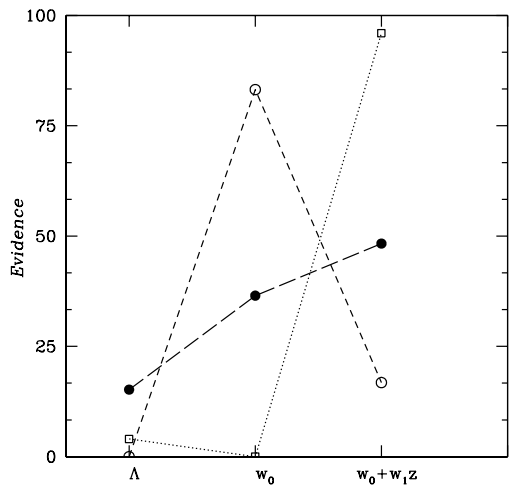,height=7cm}
\caption{Evidence values for various models, and various 
levels of fit as indicated on the x-axis, with a Gaussian prior
$\Om=0.3 \pm 0.05$, and a uniform prior on the equation of state in the
range $-1 \le w(z) \le 1$. The dotted line is for an input model with
$w(z) = -0.8+0.6z$, the short dashed line for an input model with $w(z)=-0.7$, 
the long-dashed line for a model with $w(z) = -0.8+0.3z$. 
}
\label{fig:evnew}
\end{figure}

\section{Conclusions} \label{sec:conclusion}
We have investigated the use of Bayesian evidence for establishing the
significance of fitting cosmological data to various models of the
dark energy. By focusing on a polynomial approximation for the unknown
equation of state parameter $w(z)$ we have shown how evidence can be
used to fix the polynomial order, thereby establishing the minimum
variation in $w(z)$ required to fit the data adequately.

We find that the evidence is affected by the priors. The largest
uncertainty in establishing the nature of dark energy stems from the
lack of knowledge of the precise value of the present day density in
the form of pressureless dark matter.  Our results for the currently
available SNe show that the current data favours the simplest case of
cosmological constant. Since the peak of the likelihood lies in the
range $w_0 < -1$, the tight prior $ -1 \le w(z) \le 1$ disfavours the
more complex models more strongly.

We have shown that for a SNAP like data set, if one uses ``wide''
priors on the polynomial coefficients, evidence will enable us to
decide if the data is best fit to a model with a cosmological
constant, by a constant equation of state or, if the evolution is
larger than $w_1 \sim 0.5$, to a linear model.  We have shown that our
conclusions can be significantly improved if we narrow the range of
dark energy parameters by employing a prior of the form $ -1 \le w(z)
\le 1$, which happens naturally for a subclass of dark energy
models. This prior establishes the correct order of polynomial even
for relatively modest evolution in the equation of state.  However,
since there are models which allow, albeit in a contrived way, $w<-1$,
we have allowed for these models in our examples.

To summarize, we have shown that evidence can be a powerful tool for
systematically pinning down the nature of dark energy. Its application
to the polynomial approximation for the equation of state parameter
allows us to fix the polynomial order that is required by data.  The
simple relation between the polynomial coefficients and the underlying
true equation of state enables us to obtain all the properties of dark
energy provided by the data.

\section{Acknowledgement}
We thank Rob Crittenden, Steve Gull, David MacKay, Phil Marshall,
Thanu Padmanabhan and Ben Wandelt for helpful discussions. TDS thanks
PPARC for financial support.  SLB acknowledges support from a Selwyn
College Trevelyan Research Fellowship.  JW is supported by the
Leverhulme Trust and a Kings College Trapnell Fellowship.

\newcommand{\ea}{et~al.\ }
\def\bb#1#2#3#4#5#6#7{\bibitem[\protect\citename{#2 }#3]{#1}#4, #3,
#5, #6, #7.}
\def\bbprep#1#2#3#4#5{\bibitem[\protect\citename{#2 }#3]{#1}#4, #3, #5.}
\def\prl{Phys.\ Rev.\ Lett.}
\def\pr{Phys.\ Rev.}
\def\pl{Phys.\ Lett.}
\def\np{Nucl.\ Phys.}
\def\prp{Phys.\ Rep.}
\def\rmp{Rev.\ Mod.\ Phys.}
\def\cmp{Comm.\ Math.\ Phys.}
\def\mpl{Mod.\ Phys.\ Lett.}
\def\apj{Ap.\ J.}
\def\apjl{Ap.\ J.\ Lett.}
\def\aap{Astron.\ Ap.}
\def\cqg{Class.\ Quant.\ Grav.} 
\def\grg{Gen.\ Rel.\ Grav.}
\def\mn{MNRAS}
\def\ptp{Prog.\ Theor.\ Phys.}
\def\jetp{Sov.\ Phys.\ JETP}
\def\jetpl{JETP Lett.}
\def\jmp{J.\ Math.\ Phys.}
\def\zpc{Z.\ Phys.\ C}
\def\cupress{Cambridge University Press}
\def\pup{Princeton University Press}
\def\wss{World Scientific, Singapore}
\def\oup{Oxford University Press}
\def\asj{Astron.~J}
\def\imp{Int.\ J.\ Mod.\ Phys.}
\def\ap{astro-ph/}

\end{document}